\documentclass[preprint,showpacs,preprintnumbers,amsmath,amssymb]{revtex4}

\usepackage{graphicx}
\usepackage{dcolumn}
\usepackage{bm}

\begin{document}
\title{Structure of hybrid polymer by numerical simulation: example of
 $(Si(CH_3)_2)_n(OSiO)_p(OH)_q$}

\author{   N.Olivi-Tran$^{1,2}$,S.Calas$^2$,P.Etienne$^2$}
\affiliation{$^1$S.P.C.T.S., UMR-CNRS 6638, Universite de Limoges, 47 avenue Albert Thomas
87065 Limoges cedex, France}
\affiliation{$^2$G.E.S., UMR-CNRS 5650, Universite Montpellier II, 
case courrier 074, place Eugene Bataillon, 34095 Montpellier cedex 05, France
}
\date{\today}

\begin{abstract}
 The structure depends
on the mechanism of hydrolyse and polycondensation in the case
 sol-gel process used to obtain  hybrid polymers containing uniquely
covalent bonds.
We computed by a tight binding calculation taking into account hybridization,
the total electronic energy of different structures (amorphous, fractal and linear) of $(Si-(CH_3)_2)_n-(OSiO)_p(OH)_q$. We found that 
 the total electronic energy of the amorphous structure was the smallest as a function of the number of atoms contained
by the structure. As the total electronic energy is linked to the toughness
of the structure, we may say 
that the structure of such hybrid polymer has the highest toughness
for the amorphous structure.
\end{abstract}
\pacs{36.20.Kd;31.15.bu;81.05.Lg}
\maketitle
\section{Introduction}
Hybrid materials may be classified into two families. The class I family
corresponds to hybrid materials where the organic part is embedded
in an inorganic network. The interactions between the mineral and the organic
parts are weak essentially Van der Waals, hydrogen bonds and electrostatic
interactions \cite{43}.
The class II corresponds to the existence of chemical bonds (covalent or ionic-covalent) between the organic and the mineral part of the network \cite{43}.
The synthesis of class II hybrid polymers has been initiated simultaneously
by the sol gel scientists and the polymer scientists.

The sol gel process is a method to obtain hybrid polymers: one would have
to incorporate to the sol inorganic precursors and organic compounds
with functionalities which can be plugged to the inorganic part
of the gel. This may lead to hybrid nanomaterials \cite{39}.

We deal here with class II hybrid polymers containing the species: $Si,C,O$ and $H$. This type of polymers may be obtained by the sol gel process.
In this particular type of polymers containing only covalent bonds it is possible
to use a tight binding approach to compute the total electronic energy.

Our tight binding method has been modified in order to take into account
hybridization i.e. the $\sigma$ and the $\pi$ valence electrons which
enter a covalent bond. The tight binding method depends only on the connectivity
of the atoms which enter a structure and not on the real distribution
of the atoms in space. But the calculation of the total electronic
energy allows one to compare the tenacity of different structures
like amorphous, fractal and linear molecules.
\section{Tight binding approach}
Let us remind that this is a one electron model, each electron moves in a
 mean potential  $V(r)$ which represents both the nuclei attraction and 
the repulsion of other electrons. $\sigma$ and $\pi$ electrons are separately treated :

If the molecular orbital $\sigma$ is given by:
\begin{equation}
|\Psi >= \sum_{i,J} a_{iJ} | iJ >
\end{equation}

and the energy origin taken at the vacuum level, the Hamiltonian can be written as, in the case of $sp^3$ hybridization :
\begin{equation}
H_{\sigma} = E_m \sum_{i,J} | iJ >< iJ |  +  \Delta_s\sum_{i,J,J' \neq J} | iJ >< iJ | + \beta_{\sigma}\sum_{i,i'\neq i,J}  | iJ >< iJ |
\end{equation}
($i$ and $i'$ are first neighbours) where $E_m$ is the average energy: 
$ E_m =  (E_s - 3 E_p) / (3 + 1)$ ,$E_s$ and $E_p$ are the atomic level energies,$\beta_{\sigma}$  is the usual hopping or resonance integral in H\"uckel theory (interaction between nearest neighbour atoms along the bond),
$\Delta_s$ is a promotion integral (transfer between hybrid orbitals
 on the same site) :$\Delta_s =  (E_s - E_p) / (3 + 1)$.

The Hamiltonian of the $\pi$  bonds is given by:
\begin{equation}
H_{\pi} = E_p \sum_i  | i >< i |  + \beta_{\pi}\sum_{i,i'\neq i}  | i >< i'|
\end{equation}
with $ | i >$  the $\pi$  orbital centered on atom $i$,and $\beta_{\pi}$ the hopping integral for $\pi$ levels.

 We need only 3 parameters:$\beta_{\sigma}$,$\beta_{\pi}$, and $\Delta_{\sigma}$
  for the homonuclear model which represent in fact the average potential $V(r)$ and which take into account the nuclear attraction and the dielectronic
 interactions \cite{joyes}. But due to the fact that we only take into account
 on average the nuclear and dielectronic interactions, we can only compare
 clusters with the same number of atoms.

The numerical values of the parameters are given in table 1.
\section{Results}
In figure 1, one may see the typical structure that we used for the tight
binding calculation in the case of an amorphous hybrid polymer.
The picture shows a planar molecule but this may be folded and the angles
between different atoms may not be equal to 90$^o$ and the length of the bonds 
may be changed depending on the type of atoms \cite{86,87,88,96}. Thus it represents
an amorphous structure.

Figure 2 exhibits a fractal structure of the same type of hybrid polymer.
We chose the fractal structure having a	 Cayley tree type.
Once again, we can say that the angles between atoms linked to the same 
neighbour are not generally equal to 90$^o$. Our tight binding method
uses only the connectivity and the type of atoms to calculate
the total electronic energy.

Finally, figure 3 shows a linear molecule containing the same atoms
as the two previous molecules. In the bulk, i.e. where several different
linear molecules are mixed together, this type of molecules are not straight
but may be bended.

Our tight binding calculations which takes into account hybridization
allows one to compute the total electronic energy for each of the three
preceding configurations. This is reported in figure 4 a and b.
Figure 4a is the total electronic energy as a function of the number 
of atoms in the three cases (amorphous, fractal and linear molecules).
Figure 4b is also the total electronic energy as a function of the number
of valence electrons for each type of molecules.

In figure 4a, we made a linear regression in order to obtain the slope
of the evolution of the total electronic energy.
For the amorphous molecule, the linear regression gives the following
result:
\begin{equation}
E=661.2-68.45.N_{at}
\end{equation}
where $E$ is the total electronic energy and $N_{at}$ is the number of atoms
(even of different types).
In the case of the fractal molecules, we obtain:
\begin{equation}
E=-353.6-57.08.N_{at}
\end{equation}
and in the case of the linear molecules, we obtain:
\begin{equation}
E=49.908-65.82.N_{at}
\end{equation}
All numbers are given in $eV$.

In figure 4b, which is the total electronic energy as a function of the number
of valence electrons (i.e. two times the number of valence bonds), we made also
a linear regression: the slope that we obtained was the same for the three
type of molecules i.e. $21.5 \pm 0.4 eV$ within the error bars of the computation.
\section{Discussion}
In the view of the slope given by figure 4b, i.e. the total electronic energy
as a function of the number of valence electrons, the total electronic energy
does not depend on the number of valence electrons (i.e. the number of valence
bonds multiplied by two) as its evolution is the same
for all types of molecules.
So,  for the same total number of atoms, even if the number of atoms of the same type differs for the three different
molecules that we consider, we can compare the total electronic energy as
a function of the total number of atoms for a given number of atoms.

Let us analyze equations (4),(5) and (6).
For 100  atoms of different types, the total electronic energy is the largest
for the fractal structure followed by the linear structure. The amorphous
molecules is the less stable.
But as we deal with macroscopic materials, we may extrapolate equations
(4),(5) and (6) to one mole i.e. approximatively 10$^{23}$ atoms.
In this last case, the slope of the linear regression given by figure 4a
gives the following results: the amorphous material is the most stable
then the material containing linear molecules is intermediate and finally
the fractal structure has a stability which is the lowest.For 10$^{23}$ atoms, the difference between the amorphous and the linear
materials electronic energy is approximatively equal to
$2.6.10^{23}eV=4.2.10^4J$; and the difference between the linear and the fractal
structures is equal to $8.74.10^{23}eV=13.98.10^{4}J$.

So two cases may appear: first a fractal structure of the hybrid polymer
with fractal domains not overcoming 100 atoms, and which would be more stable
than the amorphous structure and that the linear one. But a fractal domain
of 100 atoms is too small to be taken into account.
Finally we may say that the amorphous structure is the most stable
followed by the linear one.
The linear structure induces that the linear molecules are mixed up
like a plate of spaghetti.
The fractal structure may exist for less than 100 atoms then it will
be replaced by an amorphous one.

Let us remark, that our tight binding approximation deals only with connectivity,
so even if we have a fractal, linear and amorphous connectivity, the resulting
real structure may be only amorphous for all three structures.
Indeed, we did not take into account nor the length of the bonds 
neither the angle between three atoms.
\section{Conclusion}
We modelled the molecules contained in a hybrid compound such as
 $(Si(CH_3)_2)_n(OSiO)_p(OH)_q$. We calculated by a tight binding approach
modified in order to take into account the hybridization (here $sp^3$)
the total electronic energy of each type of molecules.
By making a linear regression of this total electronic energy as a function
of the number of atoms, we obtained that the amorphous molecules, for large
sizes are the most stable (i.e. have the total electronic energy the smallest).
This result may be related to mechanical properties of such material:
the toughness of the amorphous hybrid material is the largest; indeed,
the total electronic energy is related to the toughness of the valence
bonds within the structure, so the toughness of the valence bonds
can be linked to the mechanical toughness of the material.

Acknowledgements: N.O.T. would like to acknowledge M.Leleyter for helpful discussions
\pagebreak

\pagebreak
\begin{table}
\begin{center}
\begin{tabular}{|c|c|c|c|c|}
\hline
 atom/parameter & $E_{\sigma}$ & $E_{\pi}$ & $\beta_{\sigma}$ & $\beta_{\pi}$ \\
\hline
$H$ & 13.6 & 0.0 & 15.05 & 0.0 \\
\hline
$C$ & 19.45 & 10.74 & 7.03 & 3.07 \\
\hline
$O$ & 32.37 & 14.96 & 12.0 & 5.0 \\
\hline
$Si$ & 14.96 & 7.75 & 4.17 & 0.8 \\
\hline 
\end{tabular}
\end{center}
\caption{Parameters for the tight binding calculations}
\end{table}
\begin{figure}
\includegraphics[width=18cm]{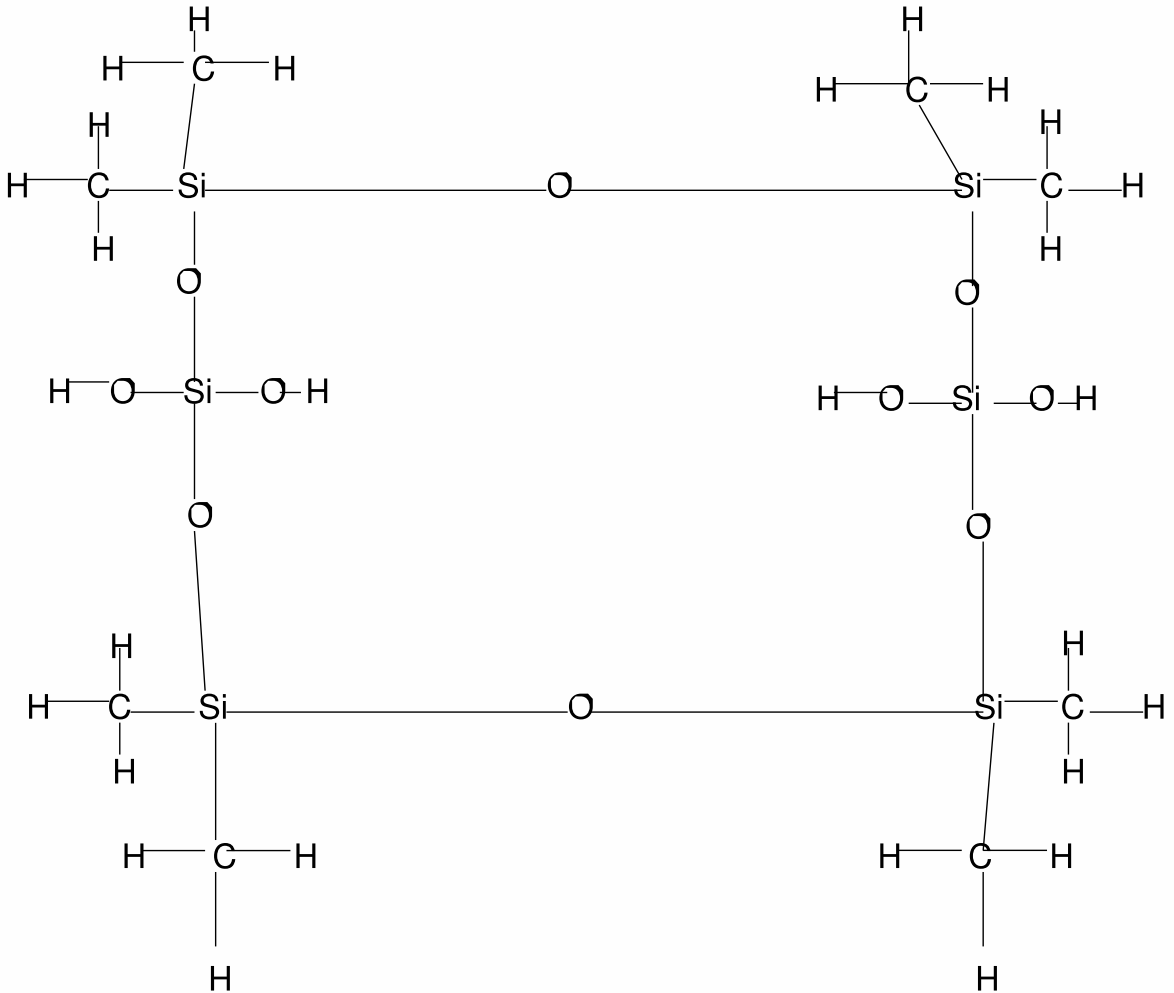}
\caption{Example of amorphous molecule}
\end{figure}
\begin{figure}
\includegraphics[width=18cm]{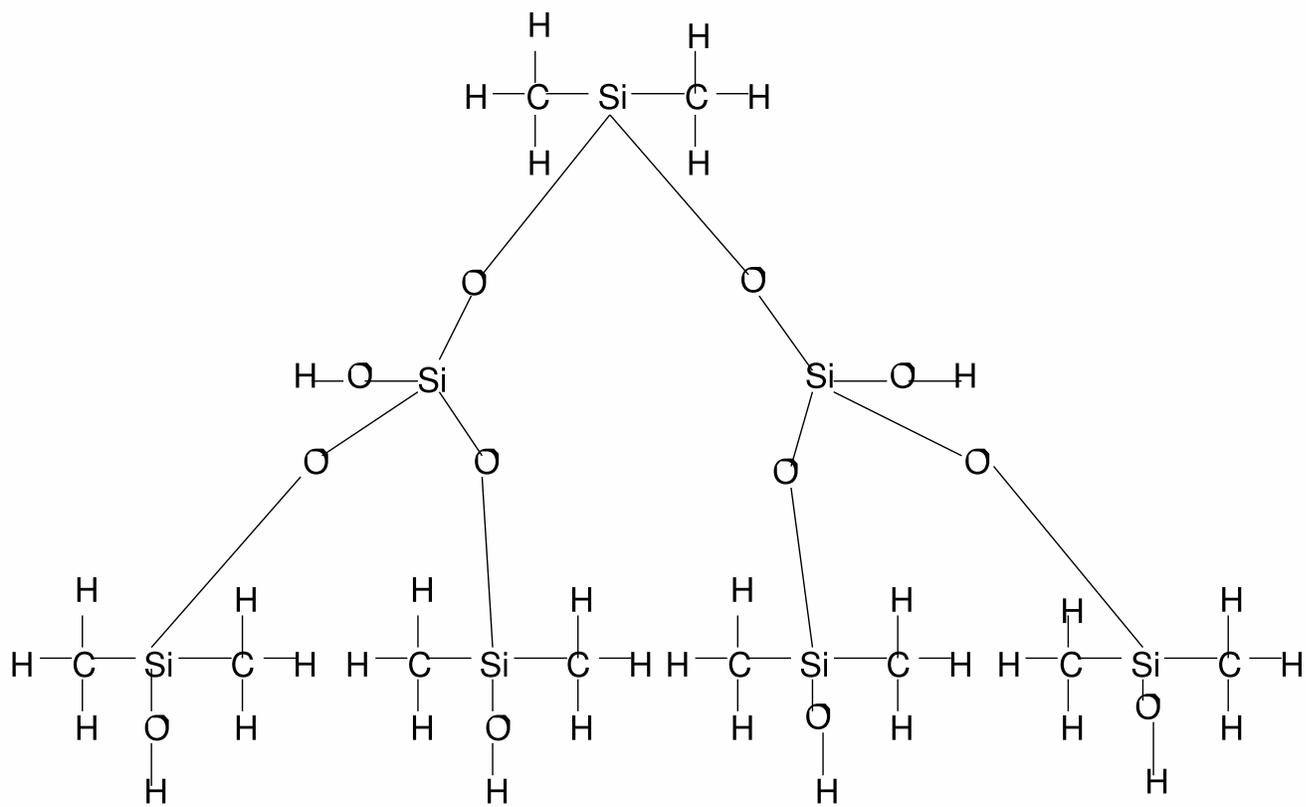}
\caption{Example of fractal molecule}
\end{figure}
\begin{figure}
\includegraphics[height=18cm]{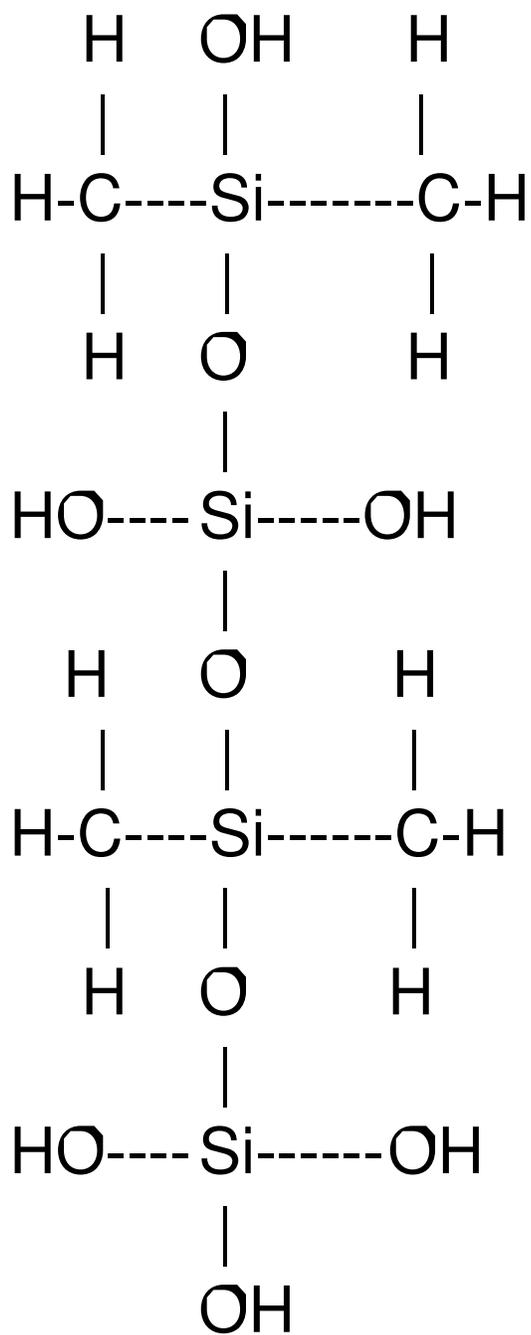}
\caption{Example of linear molecule}
\end{figure}
\begin{figure}
\includegraphics[width=18cm]{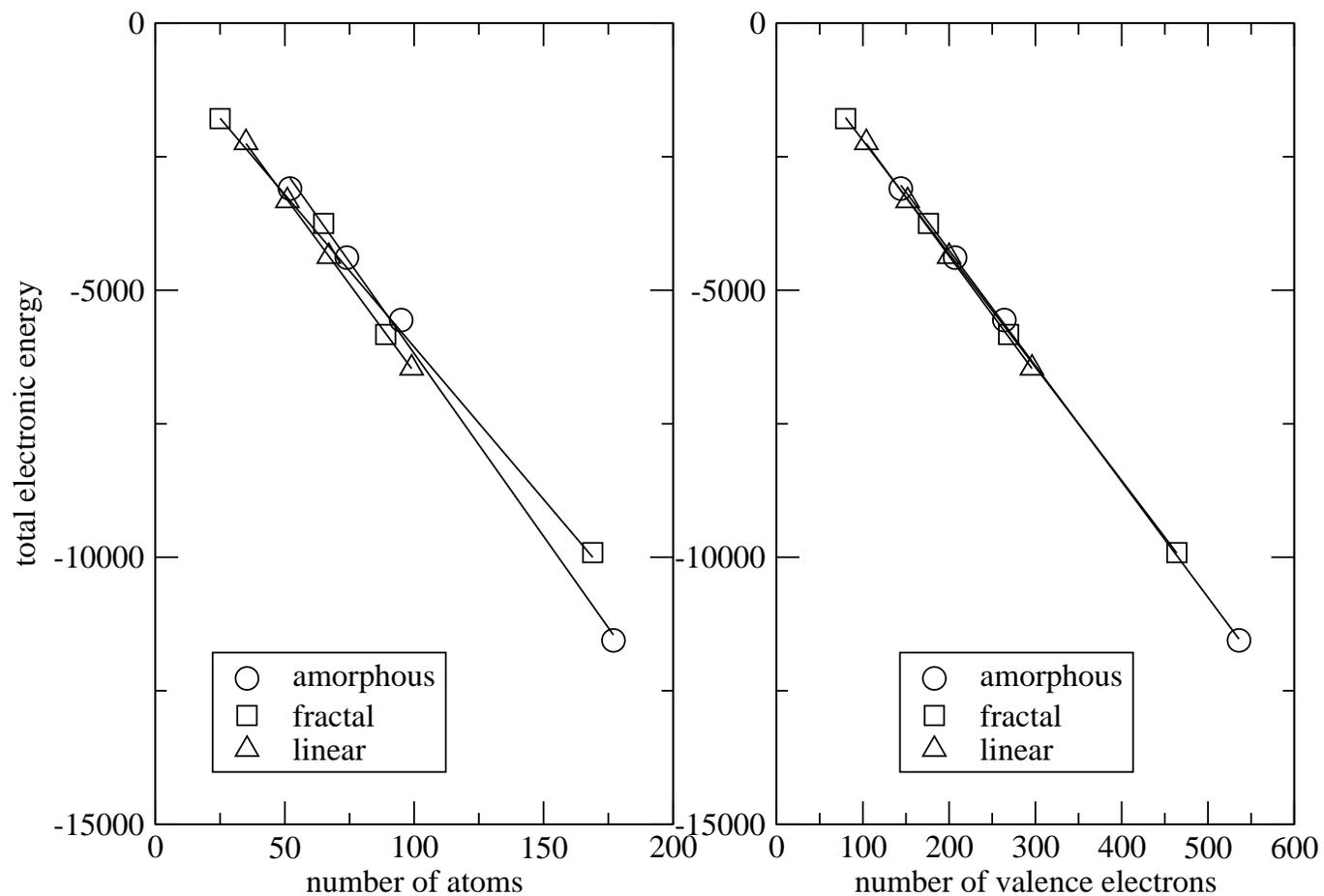}
\caption{Left: total electronic energy as a function of the number of atoms in the three cases of molecules: amorphous, fractal and linear. Right: total electronic
energy as a function of the number of valence electrons for the three cases of
molecules: amorphous, fractal and linear}
\end{figure}


\begin{thebibliography}{99}
\bibitem{43}{P.Proposito and M.Casalboni in {\it Handbook of Organic-Inorganic
Hybrid Material and Nanocomposites} {\bf 1} (2003) 83 (H.S.Nalwa Ed., American Scientific Publisher (2003))}
\bibitem{39}{C.Sanchez, D.Babonneau {\it Mat\'eriaux Hybrides} (Masson, Paris, 1996)}
\bibitem{joyes}{P.Joyes,  {\it Les Agr\'egat Inorganiques Elementaires} (Les Editions de Physique (1990) Paris) }
\bibitem{86}{H.Gunzler and H.U.Gremlich, {\it IR Spectroscopy: an Introduction}(Weinheim, Wiley-Ch, (2002))}
\bibitem{87}{D.L.Ou and A.B.Seddon, J. of Non-Cryst. Sol. {\bf 210} (1997) 187}
\bibitem{88}{N.B.Colthup,L.H.Daly and S.E.Wilberly {\it Introduction to Infrared and Raman Spectroscopy} (Academic Press (1964) New-York}
\bibitem{96}{Z.Olejniczak,M.Leczka,K.Cholewa-Kowalaska,K.Wojtach,M.Rokitaand W.
Mozgawa,J.of Mol. Struct. {\bf 744} (2005) 69}
\end{thebibliography}
\end{document}